\documentclass[pra,final,twocolumn,showpacs]{revtex4-1}%
\usepackage{graphicx,amsmath,amsfonts,wasysym,color}

\begin{document}


\title{Resistive cooling of highly charged ions in a Penning trap to a fluid-like state}

\author{M.S. Ebrahimi}
\affiliation{GSI Helmholtzzentrum f\"ur Schwerionenforschung, 64291 Darmstadt, Germany}
\author{M. Wiesel}
\affiliation{GSI Helmholtzzentrum f\"ur Schwerionenforschung, 64291 Darmstadt, Germany}
\affiliation{Institut f\"ur Angewandte Physik, Technische Universit\"at Darmstadt, 64289 Darmstadt, Germany}
\author{Z. Guo}
\affiliation{GSI Helmholtzzentrum f\"ur Schwerionenforschung, 64291 Darmstadt, Germany}
\author{G. Birkl}
\affiliation{Institut f\"ur Angewandte Physik, Technische Universit\"at Darmstadt, 64289 Darmstadt, Germany}
\author{W. Quint}
\affiliation{GSI Helmholtzzentrum f\"ur Schwerionenforschung, 64291 Darmstadt, Germany}
\affiliation{Physikalisches Institut, Universit\"at Heidelberg, 69120 Heidelberg, Germany}
\author{M. Vogel}
\affiliation{GSI Helmholtzzentrum f\"ur Schwerionenforschung, 64291 Darmstadt, Germany}

\begin{abstract}
We have performed a detailed experimental study of resistive cooling of large ensembles of highly charged ions such as Ar$^{13+}$ in a cryogenic Penning trap. Different from the measurements reported in [M. Vogel et al., Phys. Rev. A {\bf{90}}, 043412 (2014)], we observe purely exponential cooling behavior when conditions are chosen to allow collisional thermalization of the ions. We provide evidence that in this situation, resistive cooling time constants and final temperatures are independent of the initial ion energy, and that the cooling time constant of a thermalized ion ensemble is identical to the single-ion cooling time constant. For sufficiently high
ion number densities, our measurements show discontinuities in the
spectra of motional resonances which indicate a transition of the ion
ensemble to a fluid-like state when cooled to temperatures below
approximately 14\,K. With the final ion temperature presently being 7.5\,K, ions of the highest charge states are expected to form ion crystals by mere resistive cooling, in particular not requiring the use of laser cooling.
\end{abstract}

\maketitle

\section{Introduction}
Resistive cooling is a well-established technique for slowing down the motion and reducing the amplitude of oscillation of charged particles in Penning traps \cite{rs1,rs2,rs3}. It is based on the dissipation of particle kinetic energy into resistive elements that represent a heat bath at some low target temperature. To this end, a resistive circuit is set up to connect geometrically suited trap electrodes \cite{dfa}. The particle oscillation in the trap causes oscillations of the image charges induced in these electrodes (Shockley-Ramo theorem \cite{sr1,sr2}) and leads to an image current through the resistive circuit that is hence extenuated to the point where the confined charged particles are in thermal equilibrium with the circuit. Usually, a resonant (RLC) circuit is used, based on the high achievable resonance resistances which lead to efficient cooling \cite{cir1,cir2,cir3}. For highly charged ions, resistive cooling is particularly suited due to the scaling of the cooling rate with the square of the particle charge $q$ \cite{rs3}. Highly charged ions are inapt for laser cooling due to their lack of broad optical transitions \cite{nolas}, and have large charge exchange cross sections that make buffer gas cooling or electron cooling unfavourable. They can be sympathetically \cite{sym1,sym2} cooled to a crystalline state well below the ambience temperature by interaction with singly charged laser-cooled ions \cite{sym3} when storage of both species with sufficient proximity and laser access is possible.

In the present ARTEMIS experiment \cite{art0,art1}, cooling of the trapped ions is used to suppress Doppler effects in optical and microwave spectroscopy which will be used to determine the magnetic moments of the ionic nucleus and the bound electron in highly charged ions. The work presented has been performed in the framework of the HITRAP facility at GSI, Germany \cite{kluge,gfak}.
 
In the following, we will give a short account of our setup and experimental procedures, particularly for in-trap production of highly charged ions, their non-destructive detection and resistive cooling. 
Note that presently, detection and cooling are both aspects of the same interaction of the confined ions with the resistive circuit.
We will briefly review the theory of resistive cooling and of mass spectrometry of ion ensembles in a Penning trap, discuss their thermalization behavior, and how they are described in terms of a non-neutral plasma. The results section then shows spectrometric data, allowing us to determine the energy, spectral width, and spectral position of all species present in the ion ensembles during resistive cooling. While at low ion number densities, purely exponential cooling behavior in agreement with naive theory is observed, dense ion ensembles show spectral discontinuities that agree with predictions of a transition to a fluid-like state. As we present measurements with ion numbers of the order of several thousands and above, these ensembles are distinct from both the well-known single-ion case and the small ensembles studied in \cite{rs3}, and show collective behavior, depending mainly on the rate of thermalization. We compare the cooling dynamics to the non-exponential behavior reported earlier \cite{rs3} to show that fast thermalization within the ensemble is critical for efficient cooling of ion ensembles, rather than achieving high effective resistances with high cooling rates, but limited to a narrow energy distribution. This points to a potential limitation for resistive cooling in traps with highly tuned potentials or narrow-band cooling circuits.  

\section{Setup and Procedures}
\begin{figure}[h!]
\begin{center}
  \includegraphics[width=\columnwidth]{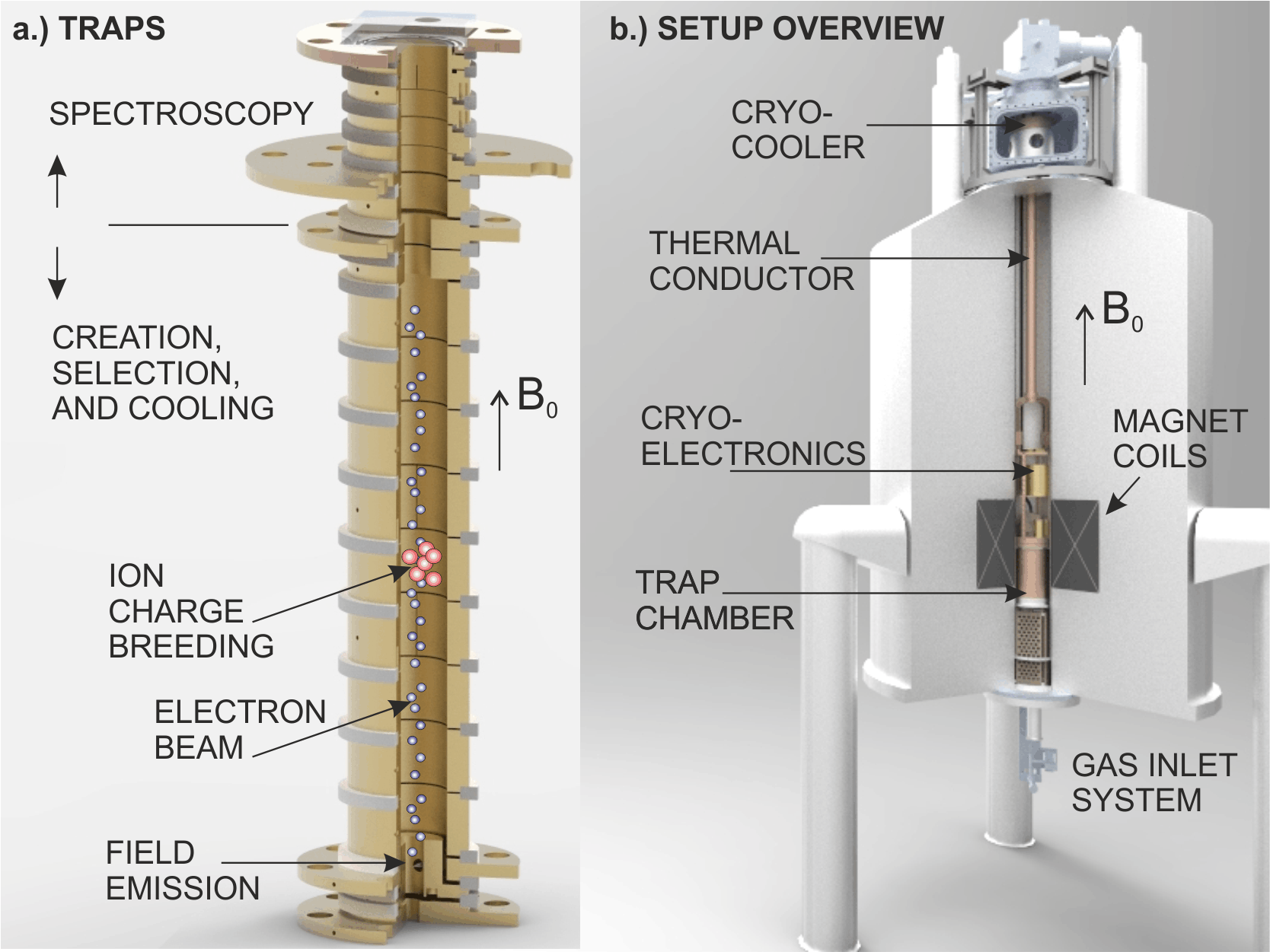}
  \caption{\small (Color online) Left: Stacked configuration of Penning traps for ion creation, selection, and cooling (lower part) and spectroscopy (upper part). Right: Overview of setup showing the cryogenic trap chamber in the field centre of the superconducting magnet.}
  \label{one}
\end{center}
\end{figure}

The experimental setup has been described in detail in \cite{art0,art1}. Briefly, an arrangement of stacked cylindrical Penning traps is located in the homogeneous field of a superconducting magnet and is cooled to liquid helium temperature (Fig. \ref{one}). This arrangement consists of a creation trap with multiple potential wells for ion confinement, and of a trap optimized for optical and microwave spectroscopy of confined ions under large solid angles \cite{ito}.

In a Penning trap, the homogeneous magnetic field $B_0$ along the $z$-direction forces the charged particles on orbits, hence confining them radially, while the electrode voltages of the trap are chosen to create a well at $z=0$ in the axial direction, thereby confining the ions to a small volume around the trap center $\rho=z=0$. In the absence of ion-ion interactions, each individual ion performs an oscillatory motion consisting of three eigenmotions, two in the radial plane perpendicular to the magnetic field, and one axial oscillation parallel to the trap axis \cite{bro86}. 

In a cloud of interacting ions, the overall motion is more complicated on account of ion-ion interactions. Nevertheless, the axial and radial oscillatory motions in arbitrary ion clouds can be detected non-destructively, which is used to perform charge-to-mass spectrometry of the stored particles. At present, we are concerned with the detection and cooling of the axial motion only.
In the creation trap, ensembles of highly charged ions are produced by impact ionization of gas by an electron beam from a field emission tip, in close similarity to the charge-breeding process in electron-beam ion sources \cite{art1,joseba}. 

\begin{figure}[h!]
\begin{center}
  \includegraphics[width=\columnwidth]{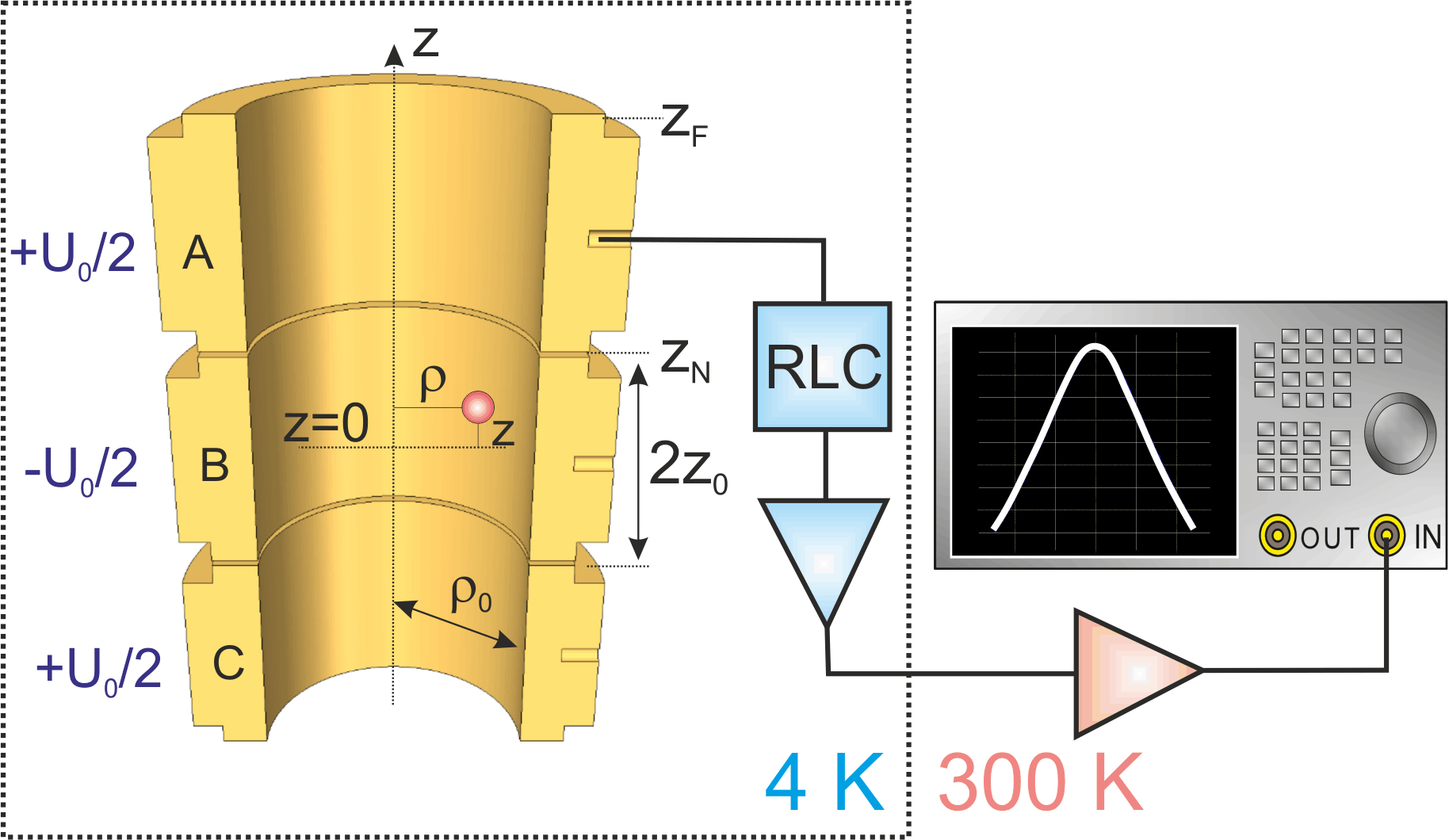}
  \caption{\small (Color online) Schematic of the part of the creation trap used for the cooling measurements (cut for presentation of the inner structure): electrode A is connected to a resonant (RLC) circuit used for both detection and cooling of the axial ion motion and is read out by a spectrum analyzer via two amplifiers. All elements within the dotted box are cooled to liquid-helium temperature.}
  \label{two}
\end{center}
\end{figure}

Figure \ref{two} shows the part of the creation trap relevant for ion detection and the cooling measurements: Three electrodes A, B, and C form a geometrically compensated trap \cite{gab84} with B acting as the ring electrode and A and C being the endcaps. Electrode A is connected to a resonant (RLC) circuit for detection of the axial ion motion. The trap voltage $U_0$ is symmetrically split between ring and endcap electrodes as indicated.

\begin{figure}[h!]
\begin{center}
  \includegraphics[width=\columnwidth]{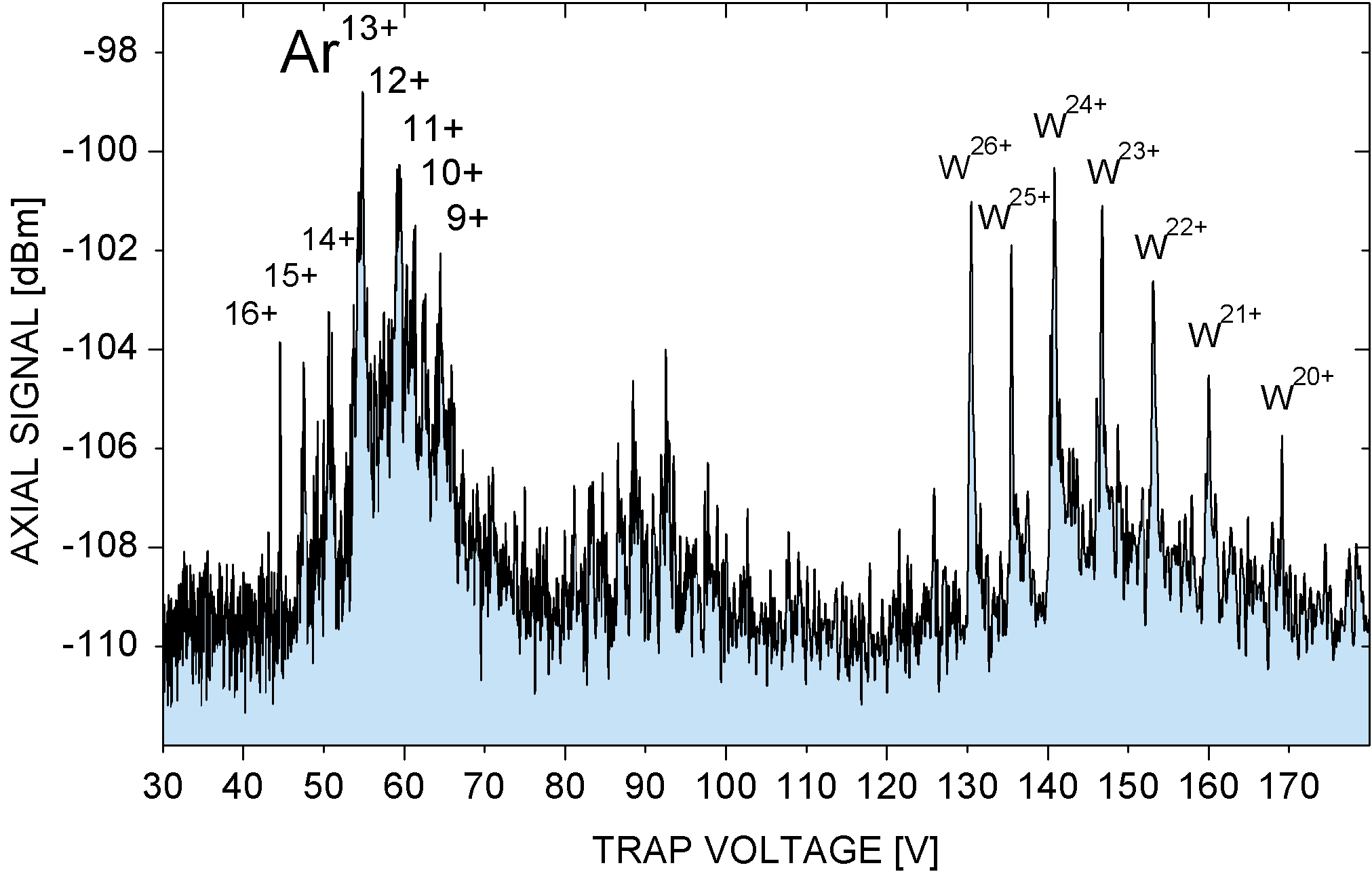}
  \caption{\small Ion spectrum upon ion creation. It shows charge states 9+ to 16+ of the desired argon ions and a number of charge states of tungsten from the field emission tip.}
  \label{three}
\end{center}
\end{figure}

Non-destructive spectrometry of the confined ions is performed by recording the signal induced in the axial resonator circuit used for cooling as a function of the trap voltage (Fig. \ref{three}). Presently, the detection system is resonant at a fixed frequency of $\omega_R=2\pi\times 741.23$\,kHz and has a quality factor of about $Q=900$, such that the spectral width of the circuit is $\delta \omega_R=\omega_R/Q \approx 2 \pi \times 820$\,Hz.
When the trap voltage $U_0$ is scanned, ion species with different charge-to-mass ratios $q/m$ subsequently come into resonance when their axial oscillation frequency $\omega_z(U_0)$ (see Eq. (\ref{z})) matches the resonance frequency $\omega_R$ of the detection system, i.e. when for their specific value of $q/m$ we have $\omega_z(U_0)=\omega_R$. Since the detection system records a spectral power density within its frequency
band for a certain scan time, each peak area is proportional to an energy. To be more
precise, it is proportional to the individual particle kinetic energy times the number of particles contributing to that peak. This means that at constant particle number, the area measures the individual particle energy.

In our experiment, the particle number can indeed be regarded as constant, since ion loss due to collision or charge exchange with neutral species has been observed to be negligible on the time scale of the cooling experiments. As expected from kinetic gas theory, the cryogenic temperature of about 4\,K ensures efficient cryo-pumping of residual gases. From a measurement of the ion signal as a function of time over the course of 48 hours, a charge state lifetime (half-life) for Ar$^{13+}$ of 22 days is extracted, which at a cross section for charge exchange with helium gas of $(3.25 \pm 0.25) \times 10^{-15}$\,cm$^2$ \cite{mann} indicates a residual gas pressure of about $3 \times 10^{-16}$\,hPa.

Figure \ref{three} shows a spectrum of ions produced from supplied argon gas which also contains tungsten ions sputtered from the field emission tip. The signals in-between could not be assigned to an ion species. The spectrum is cleaned from all ions except argon prior to the following measurements by application of the SWIFT technique \cite{li} to the axial motion, i.e. by simultaneous resonant ejection of all unwanted species from the trap. This selective excitation increases the axial ion kinetic energy of unwanted species until it exceeds the trap depth and the ions are axially lost from confinement. The procedure takes place on the milli-second time scale and is presently combined with a temporal lowering of the trap potential to ease the ejection. Depending on the details, the number of unwanted ions is lowered to undetectable levels, while spurious excitation reduces the number of wanted ions by up to about 50\,\%. The result of this procedure (upon further cooling) is a spectrum as shown in Fig. \ref{peaks}. 
The argon ions are confined in a specific well of the creation trap for the present experiments, see Fig. \ref{two}. 
In full similarity to the cleaning procedure above, pure ion clouds containing only a single ion species even in a selected $q/m$-state can be produced by resonant ejection of unwanted ions from the trap. This is done by resonant dipole excitation of the axial oscillation again using the SWIFT technique \cite{li}.

\section{Theory of Resistive Cooling}
In an ideal cylindrical Penning trap, a single confined ion 
obeys the axial equation of motion \cite{gho}
\begin{equation}
\label{eins}
\frac{\mbox{d}^2}{\mbox{d}t^2} z + \omega_z^2 z  =0,
\end{equation}
where the axial oscillation frequency $\omega_z$ follows from the axial trapping potential 
\begin{equation}
\label{harm}
V(z)=\frac{U_0C_2}{2d^2} z^2
\end{equation}
according to
\begin{equation}
\label{pott}
\omega_z^2 z = \frac{q}{m} \frac{\mbox{d}V}{\mbox{d} z},
\end{equation}
such that for the present geometry the frequency of axial oscillation $\omega_z$ is given by \cite{gho}
\begin{equation}
\label{z}
\omega_z=\sqrt{\frac{qU_0C_2}{md^2}} \;\;\; \mbox{with} \;\;\, d^2=\frac{z_0^2}{2}+\frac{\rho_0^2}{4}.
\end{equation} 
Here, 
$C_2U_0$ constitutes the depth of the potential well for axial confinement, $z_0$ and $\rho_0$ are the axial and radial dimensions of the trap leading to $d=6.73$\,mm, and $C_2=0.563$ is a coefficient that accounts for our cylindrical trap geometry \cite{bro86,gab89}.

Within this description, resistive cooling may be modelled as a friction force which depends on the axial ion velocity {d$z$/d$t$}. The equation of motion then reads \cite{gho}
\begin{equation}
\label{vier}
\frac{\mbox{d}^2}{\mbox{d}t^2} z -  \gamma \frac{\mbox{d}}{\mbox{d}t}z +\omega_z^2 z =0,
\end{equation}
where $\gamma$ denotes the cooling rate. 
It is given by \cite{book}
\begin{equation}
\label{cool}
\gamma=\frac{R}{D^2} \frac{q^2}{m},
\end{equation}
where $R$ is the resonance resistance of the circuit used for cooling and $D$ is the so-called effective electrode distance which will be detailed below. The evolution of the ion kinetic energy in the presence of resistive cooling is given by
an exponential decay of the kind \cite{gho}
\begin{equation}
\label{dec}
E=E_0 \exp\left(-\gamma t \right).
\end{equation}
Note, that the cooling rate $\gamma$ and the exponential shape of the cooling curve (energy as a function of time) according to equations (\ref{cool}) and (\ref{dec}) is predicted to be independent of time and of the absolute value of the energy $E_0$ at the beginning of the cooling. 

Further, it is an important prediction of a detailed calculation that for a thermal ensemble of identical particles, the cooling rate of the total kinetic energy is expected to be identical to the single-particle case \cite{rs3,gho,book}.

The effective electrode distance $D$ in equation (\ref{cool}) contains the information about the location and the geometry of the electrodes connected by the circuit with respect to the center of the ion oscillation \cite{book}. For the present geometry (see figure \ref{two}) with electrode (A) used for pick-up of the signal while the trap center is in electrode (B), $D$ can be calculated by \cite{book}
\begin{equation}\label{effd}
D^{-1}(\rho,z)=\frac{\partial}{\partial z} \Xi(\rho,z).
\end{equation}
where $\Xi(\rho,z)$ reads
\begin{equation}\nonumber
\Xi(\rho,z)=\frac{-1}{\rho_{0}\pi}\int_{0 }^{\infty}\frac{\textrm{I}_0~(x\rho/\rho_0)}{\textrm{I}_0~(x)} G \mbox{d}x,
\end{equation}
in which the geometry function $G$ is given by
\begin{equation}\nonumber
G=\textrm{sinc}(x\frac{z-z_F}{\pi \rho_0})(z-z_F)- \textrm{sinc}(x\frac{z-z_N}{\pi \rho_0})(z-z_N)
\end{equation}
and $\rho_0$ is the inner radius of the cylindrical electrodes, $z_N$ and $z_F$ are the distances from the center of the trap to the nearest and farthest edge of the 
\begin{figure}[h!]
\begin{center}
  \includegraphics[width=0.9\columnwidth]{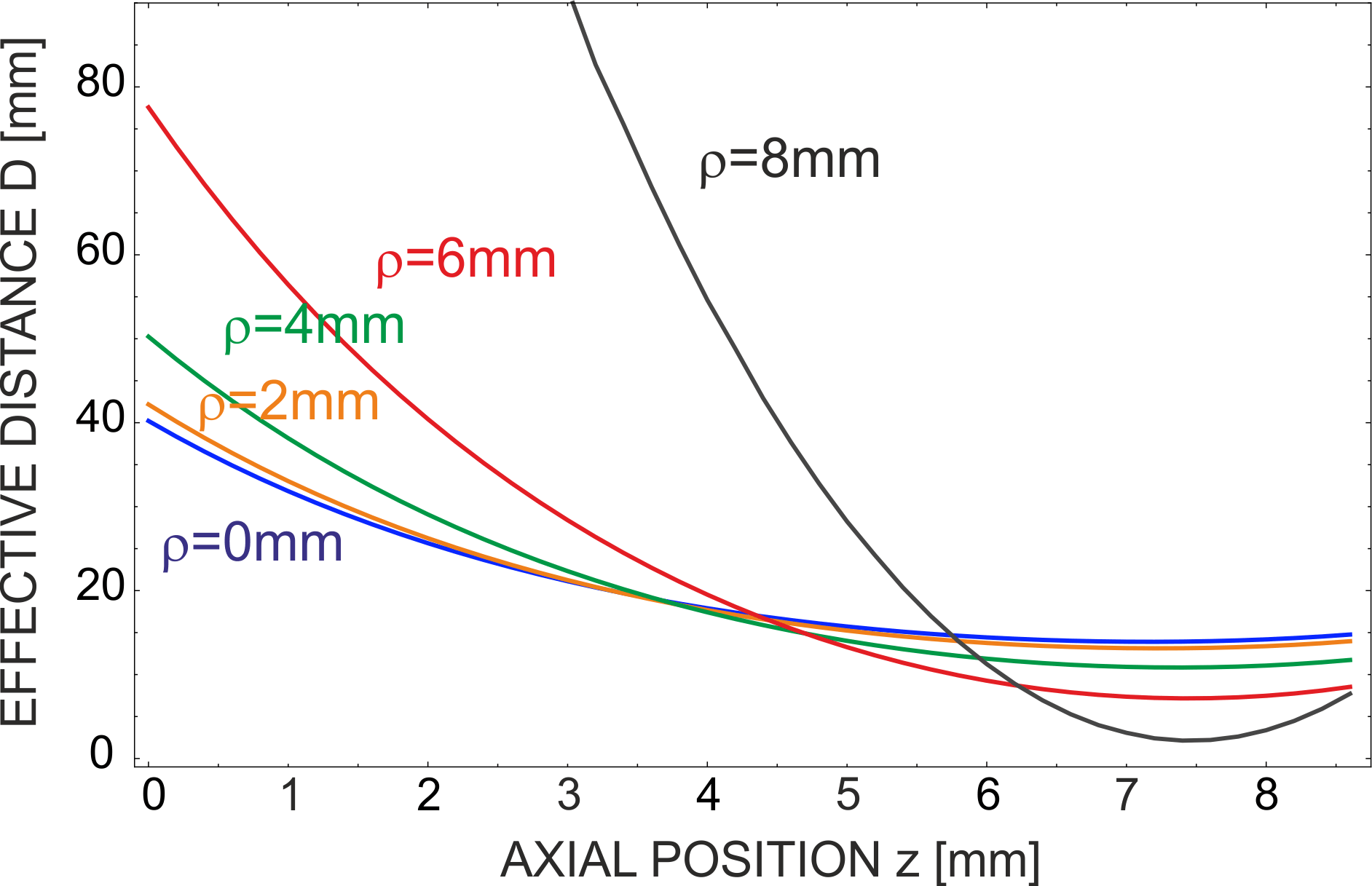}
  \caption{\small (Color online) Plot of the effective electrode distance $D$ according to Eq. (\ref{effd}) as a function of the axial position $z$ for different values of the radial coordinate $\rho$ for the geometry shown in Fig. \ref{two}.}
  \label{dee}
\end{center}
\end{figure}
electrode used for cooling (against ground), and $\textrm{I}_0(x)$ is the modified Bessel function of first kind. In our case, we have geometry values of $\rho_0 = 8.70$\,mm, $z_N= 7.46$\,mm and $z_F = 21.57$\,mm. Figure \ref{dee} shows the values of $D$ resulting from equation (\ref{effd}) as a function of the axial position $z$ for different radial positions between the trap centre $\rho=0$ and the inner electrode surface $\rho=\rho_0$. Obviously, the value of $D$ depends on both the axial and radial position in the trap, such that the expected cooling rate $\gamma$ (and likewise the cooling time constant $\tau=1/\gamma$) also depends on the ion location. Hence, for a large and/or hot ion ensemble that occupies a significant volume around the trap centre, this effect needs to be taken into account and the cooling rate has to determined via integration over the full ion distribution.

\section{Ion Ensemble Properties}

\subsection{Ion Number and Space-Charge Effects}
\label{spacecharge}
The ion number density in the trap can be deduced from the observed space-charge shift of the axial oscillation frequency of a given ion species upon production and initial cooling. 
The presence of the other charges effectively reduces the trap potential experienced by an ion and thus reduces its axial frequency to a value \cite{yu}
\begin{equation}
\omega_z'=\omega_z \sqrt{1-\frac{\omega_p^2}{3 \omega_z^2}},
\end{equation}
where $\omega_p^2=q^2n/(\epsilon_0m)$ is the plasma frequency that depends mainly on the ion charge $q$ and ion number density $n$.
In spectrometry, the space-charge shift is seen as a shift of the observed trap voltage $U'_0$ that brings a given ion species into resonance with the detection circuit relative to the single-ion value $U_0$. From this shift, the ion number density can be calculated as
\begin{equation}
n= \frac{3\epsilon_0m}{q^2} \omega_R^2 \left[1-\left( \frac{U'_0}{U_0} \right) \right].
\end{equation}
In the present measurements, $n$ is between a few $10^3$ cm$^{-3}$ and a few $10^6$ cm$^{-3}$, depending on the values of the creation parameters (trap depth, charge breeding duration, electron current and gas density).
Reproduction of a specific ion number density is difficult, a given value can be measured by the voltage shift to a factor of less than two.
At the present trap parameters, the density is hence up to 10\,\% of the electric trap limit (from the stability criterion $\omega_c^2>2\omega_z^2$ with $\omega_c=qB_0/m$ \cite{book}) and up to 1\,\% of the Brillouin limit \cite{plas,book} at the magnetic field strength of 7\,T. The observed corresponding shifts are non-negligible and care is required for a proper identification of the ion species in the spectrum.

\subsection{Spectral Properties as a Function of Energy}
\label{axtrans}
\begin{figure}[h!]
\begin{center}
  \includegraphics[width=\columnwidth]{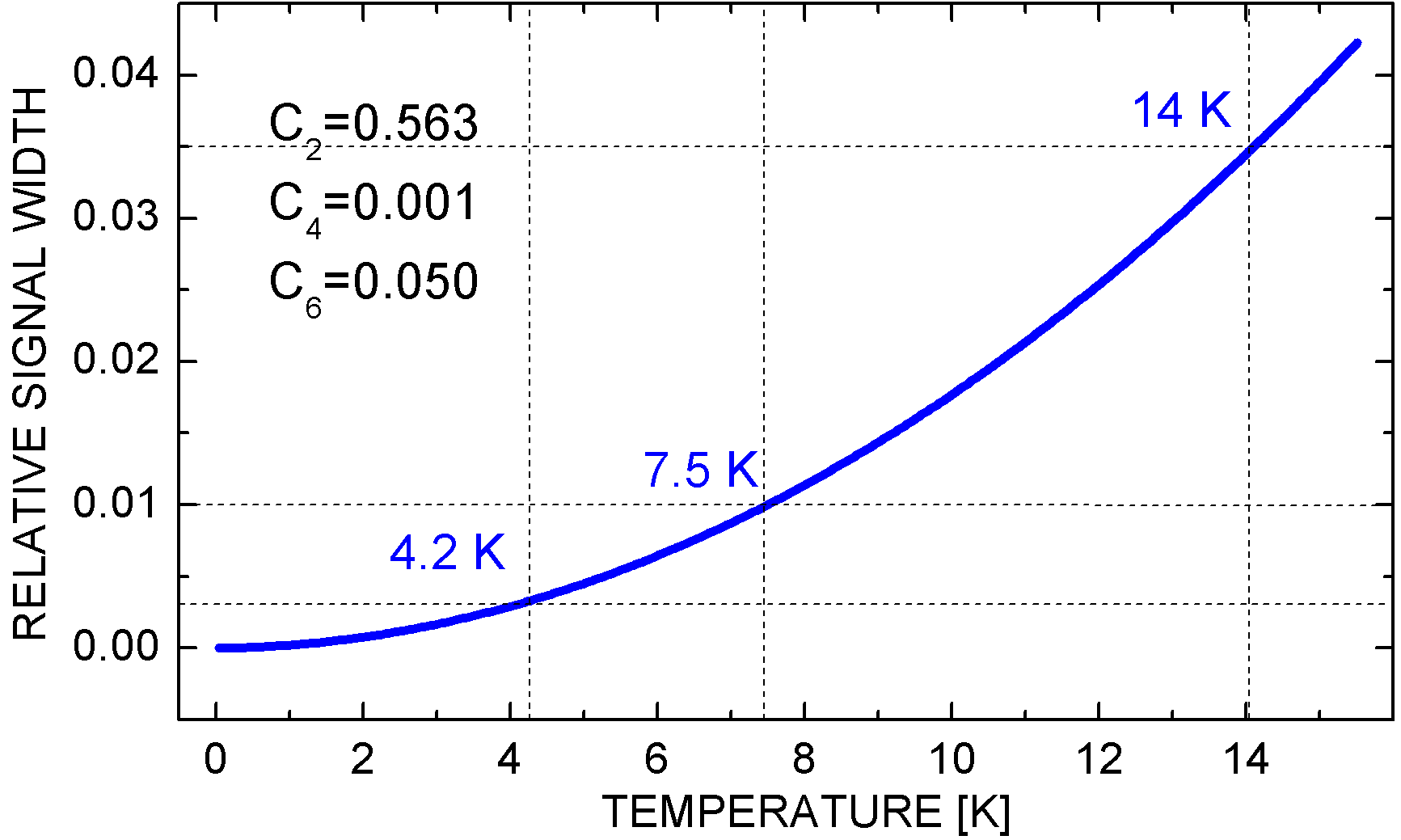}
  \caption{\small (Color online) Relative frequency width $\delta \omega_z/\omega_z$ of the axial ion oscillation as a function of the ion ensemble temperature $T$, according to Eq. (\ref{wid}).}
  \label{width}
\end{center}
\end{figure}

Imperfections of the confining electric or magnetic fields make the axial oscillation frequency of any given ion dependent on its kinetic energy. 
The most relevant imperfections of the confining fields are deviations of the electric field from the quadrupolar case. These effects have been carefully discussed in \cite{bro86,gab89,sens}. In traps like the present one, the dominant 
contributions to an energy-dependent shift of the axial frequency come from higher-order dependences of the axial trapping potential on the coordinates and are quantified by coefficients $C_4$ and $C_6$ as defined in \cite{bro86,gab89}. The relative shift of the axial oscillation frequency of any given ion with its axial energy $E_z$ is given by \cite{gab89}
\begin{equation}
\label{shift}
\frac{\Delta \omega_z}{\omega_z} = \frac{3}{2} \frac{C_4}{C_2^2} \frac{E_z}{qU_0} + \frac{15}{4} \frac{C_6}{C_2^3} \left( \frac{E_z}{qU_0} \right)^2 .
\end{equation}
Hence, for a thermal distribution of ion energies we expect a corresponding distribution of axial oscillation frequencies, shifting and broadening the signal of each ion species in the spectrum. For a thermalized ion cloud at temperature $T$, the distribution of axial and radial energies is Boltzmann-like, with the expectation value $\langle E_z \rangle = k_BT/2$ and a typical width of roughly $2k_BT$ \cite{rs3}.
Hence, we expect the average relative shift of the axial oscillation frequency distribution of each ion species (with respect to $T$=0) to be given by
\begin{equation}
\label{shift1}
\langle \frac{\Delta \omega_z}{\omega_z} \rangle = \frac{3}{4} \frac{C_4}{C_2^2} \frac{k_BT}{qU_0} + \frac{15}{16} \frac{C_6}{C_2^3} \left( \frac{k_BT}{qU_0} \right)^2 ,
\end{equation}
and the relative width of the axial oscillation frequency distribution of each ion species to be given by
\begin{equation}
\label{wid}
\frac{\delta \omega_z}{\omega_z} \approx \frac{3C_4}{C_2^2} \frac{k_BT}{qU_0} + \frac{15C_6}{C_2^3} \left( \frac{k_BT}{qU_0} \right)^2 .
\end{equation}
In principle, the ensemble temperature $T$ can be determined either from an observed signal shift or width when the coefficients $C_i$ are known, however the determination via the shift requires zero-point information and also can be obstructed by space-charge effects. Thus, a determination via the observed signal width is preferred. In the present experiment, we have $C_2 = 0.563$, $C_4 = 0.001$, and $C_6 = 0.05$. These coefficients have been calculated from the measured trap geometry according to the formalism in \cite{gab89}.
The corresponding dependence of the relative signal width on the ensemble temperature is shown in Fig. \ref{width} for a trap potential chosen such that the ion species are in resonance with the circuit at $\omega_R=2\pi\times 741.23$\,kHz. Note that in a thermal ensemble, the relative widths of the signals as a function of temperature are identical for all ion species, i.e. identical for all ion charge states present in a spectrum.

\subsection{Collisional Thermalization}
\label{rad}
Assuming an ion cloud with an arbitrary initial energy distribution and in the absence of external forces, ion-ion interactions (Coulomb collisions) thermalize the ions, eventually leading to the same Boltzmann distribution of energies within each degree of freedom. 
To quantify the time scale for this, we use the thermalization time constant ('Spitzer self-collision time'), estimated by \cite{plas}
\begin{equation}
\label{dav}
\tau_T \approx (4\pi\epsilon_0)^2\, \frac{3\sqrt{m}\, (k_BT)^{3/2}}{4 \sqrt{\pi}\, n\,q^4 \ln \Lambda},
\end{equation}
where $\ln \Lambda$ is the so-called 'Coulomb logarithm' given by \cite{plas}
\begin{equation}
\ln \Lambda =  23- \ln \left(\frac{2nq^6}{e^6T^3} \right)^{1/2},
\end{equation}
where the ion number density $n$ is given in units of cm$^{-3}$ and $T$ is given in units of eV.
Assuming argon ions of an average charge state of 11 (which is the average squared charge state of the ion distribution), even at high ion energies upon production of 10 eV and low densities of order $10^3$ cm$^{-3}$, we have $\ln \Lambda \approx 15$, and the thermalization time constant is below one second. This is much smaller than the scan time of 550\,s for recording of an ion spectrum such as Fig. \ref{three} and also smaller than the time of each voltage step applied during the scan. Additionally, during cooling, as the density increases, the thermalization time constant becomes successively smaller. From this, we deduce that in all present measurements, the ion ensembles are thermalized during cooling and spectrometry.

\subsection{Ion Ensemble as a Plasma}
An ensemble of ions confined in a Penning trap can be characterized in terms of a non-neutral plasma \cite{dub}. Depending on the strength of the ion-ion interaction via the Coulomb force in relation to the ion kinetic energy, the collective behavior of the ion ensemble can be gas-like, fluid-like, or crystal-like. The strength of the motional correlation is usually expressed by the so-called `plasma parameter' \cite{book}
\begin{equation}
\label{gam}
\Gamma=\frac{q^2}{4\pi\epsilon_0ak_BT},
\end{equation}
where $a^3=3/(4\pi n)$. We have a weakly correlated plasma (gas-like state, independent-ion picture) for $\Gamma \ll 1$, an intermediate correlation (fluid-like state) between $\Gamma \approx 2$ and $\Gamma \approx 170$, and strong correlation (ions crystals) for about $\Gamma > 170$ \cite{dub99}. Due to the factor $q^2$ in the numerator, highly charged ions have much higher degrees of correlation at the same given temperature when compared to singly charged ions. In turn, they reach fluid-like or solid-like states already at higher temperatures. 

An ion fluid can be seen as a distinct state between the largely uncorrelated ion gas and the rigid ion crystal. 
In an ion crystal, a geometric ordering results from the mutual repulsion of the ions within the common potential well of the trap, and the dominance of the Coulomb interaction energy between ions over their kinetic energy leads to a high degree of motional correlation, with each ion being bound to a fixed location within the geometric structure \cite{drew}.
Ions in a fluid perform larger excursions within a certain volume around these locations on account of their higher kinetic ion energy, and may thermally `hop' from one such volume to another one. Correspondingly, in an ion fluid, the amplitude of motion of each individual ion is comparable to the average ion-ion distance, whereas in an ion crystal, it is at least one order of magnitude smaller. Still, the degree of motional correlation is too high for independent-particle pictures to be applicable, as the Coulomb interaction energy is at least comparable to the ion kinetic energy. Possible scientific applications of such highly correlated ion systems have been discussed in \cite{fifteen}.

\begin{figure}[h!]
\begin{center}
  \includegraphics[width=\columnwidth]{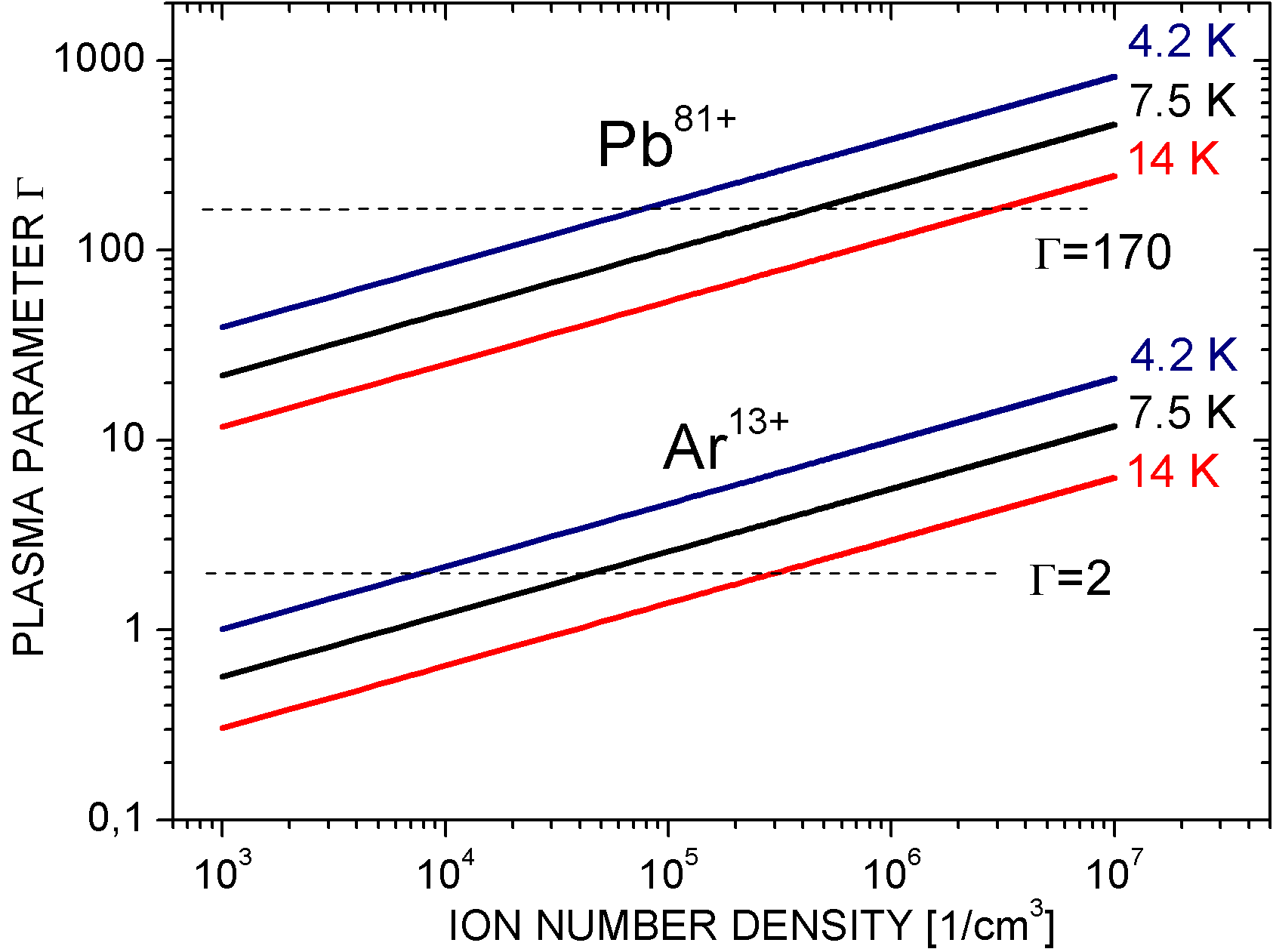}
  \caption{\small (Color online) Plasma parameter $\Gamma$ as a function of the ion number density according to equation (\ref{gam}) for Ar$^{13+}$ at three different values of the ensemble temperature $T$. For comparison, the same is shown for Pb$^{81+}$ ions.}
  \label{gamma}
\end{center}
\end{figure}
Figure \ref{gamma} shows the plasma parameter $\Gamma$ according to Eq. (\ref{gam}) for Ar$^{13+}$ at temperatures of 4.2\,K, 7.5\,K, and 14\,K as a function of the ion number density $n$. The dotted lines indicate $\Gamma=2$ and $\Gamma=170$ above which a fluid-like and crystal-like behavior of the ion ensemble is expected, respectively. Accordingly, for our conditions with ion number densities in the range of a few $10^3$ cm$^{-3}$ to a few $10^6$ cm$^{-3}$ and temperatures below 15\,K the regime of intermediate (fluid-like) correlations should be accessible. 
For comparison, $\Gamma$ is shown also for Pb$^{81+}$ ions. Here, the same plasma parameter $\Gamma$ is reached already at higher temperatures $T$ and even the crystal-like state might be achieved solely by resistive cooling.

\section{Results}
\subsection{Cooling Effect on the Ion Energy}
\begin{figure}[h!]
\begin{center}
  \includegraphics[width=\columnwidth]{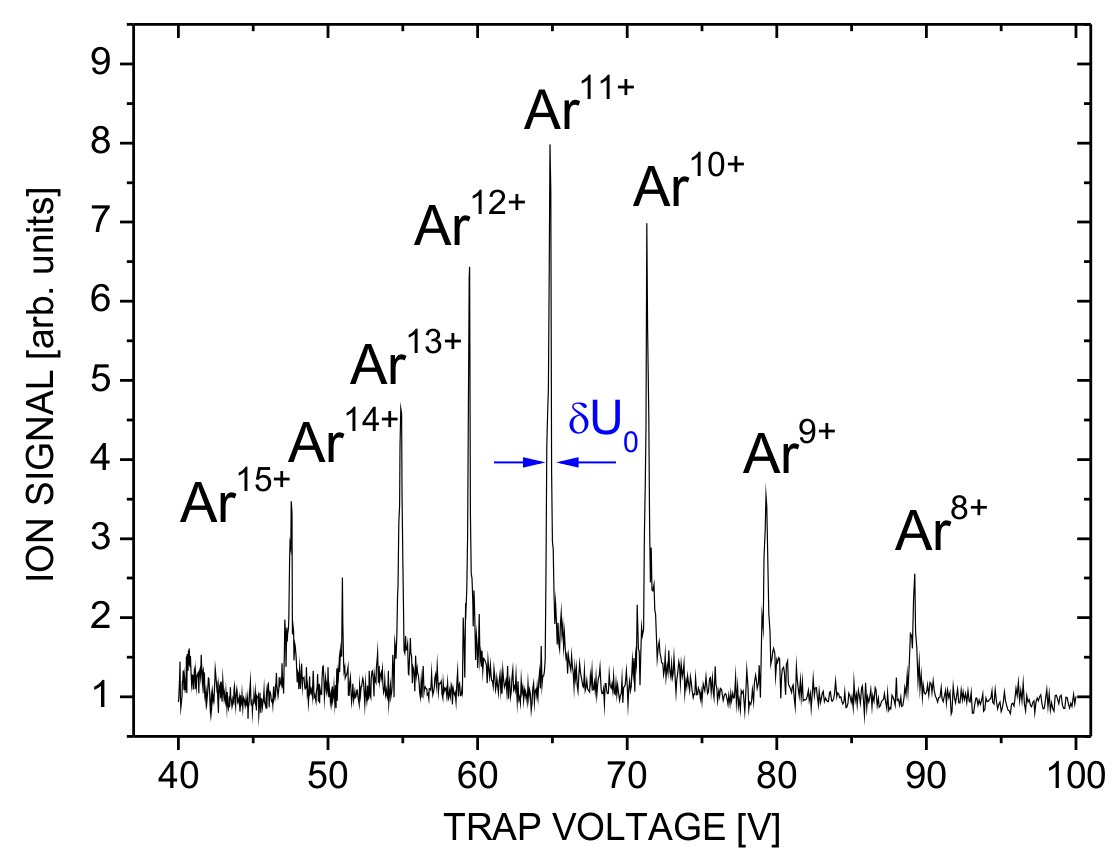}
  \caption{\small (Color online) Ion signal as a function of trap voltage. This is a typical spectrum obtained from a voltage scan (`cooling cycle').}
  \label{peaks}
\end{center}
\end{figure}

Argon ions have been created, confined, and cooled in the creation trap segment of the stacked Penning trap setup shown in Fig. \ref{one}. 
Cooling is achieved by scanning the trap voltage and hence subsequently bringing each part of the ion spectrum into resonance with the cooling circuit, detecting the spectrum at the same time, see also Fig. \ref{two}. Each individual scan ('cooling cycle') takes 550\,s and is repeated to monitor the temporal evolution of the spectrum. So each data point of the resulting cooling curve is the outcome of one such scan.
Figure \ref{peaks} shows a typical spectrum.
The full width at half maximum of a peak is indicated by $\delta U_0$, see for example the peak of Ar$^{11+}$ in Fig.\,\ref{peaks}. The value of $\delta U_0$ yields the width $\delta \omega_z$ of the corresponding axial frequency distribution according to
\begin{equation}
\delta \omega_z=\left( \frac{q}{md^2}  \right)^{1/2} U_0^{-1/2} \, \delta U_0.
\end{equation}

For the present measurements, we have chosen the timing conditions such that thermalization amongst all ion species is much faster than resistive cooling, so that cooling affects all ion species similarly at all times. The time constant for thermalization $\tau_T$ (see Eq. \ref{dav}) is of order seconds and below, which is much smaller than the time of 550\,s used for each individual voltage scan. This is done to investigate whether the non-exponential cooling behaviour of ensembles of highly charged ions reported earlier \cite{rs3} can be attributed to insufficient thermalization. 

In order to bring the present measurements of cooling by subsequently putting all species in the ion spectrum in resonance with the fixed circuit in accordance to the model of cooling underlying Eq. (\ref{cool}) (which assumes that any ion is in resonance with the circuit all of the time), we introduce two correction factors: One accounts for the fraction of the resonator bandwidth with respect to the total frequency span of the spectrum.
The frequency span $\delta$ of the spectrum is roughly $2 \pi \times 330$\,kHz, of which the resonator covers $\delta \omega_R \approx 2 \pi \times 820$\,Hz at a time. Hence, the ratio $\delta \omega_R/\delta \approx 1/400$ accounts for how much of the complete spectrum is covered by the resonator at any time.
The second correction factor concerns the shape of the spectrum, in specific the fact that the spectrum consists of discrete resonance lines. Cooling is only effective within a resonance line but not between the lines. 
To compare this to a situation in which all ions are in resonance all of the time, the summed widths of the resonance lines with respect to the full span of the spectrum has to be accounted for. When we use the sum $p$ of all peak widths and the total spectral span $\delta$, the ratio $p/\delta$ is a reasonable correction factor accounting for the spectral shape. 
The equivalent duration $t_{eq}$ of a cooling cycle is then given by the product of the cycle time $t_{cyc}$ and the two correction factors,
\begin{equation}
t_{eq}= \frac{p}{\delta} \frac{\delta \omega_R}{\delta} t_{cyc} =\frac{p \delta \omega_R}{\delta^2} t_{cyc}.
\end{equation}
and the effective time $t$ after a certain number of cycles is just the sum of all $t_{eq}$ summed up to this point. From here on, the time $t$ used in the discussions is the effective time obtained with this procedure.

\begin{figure}[h!]
\begin{center}
  \includegraphics[width=\columnwidth]{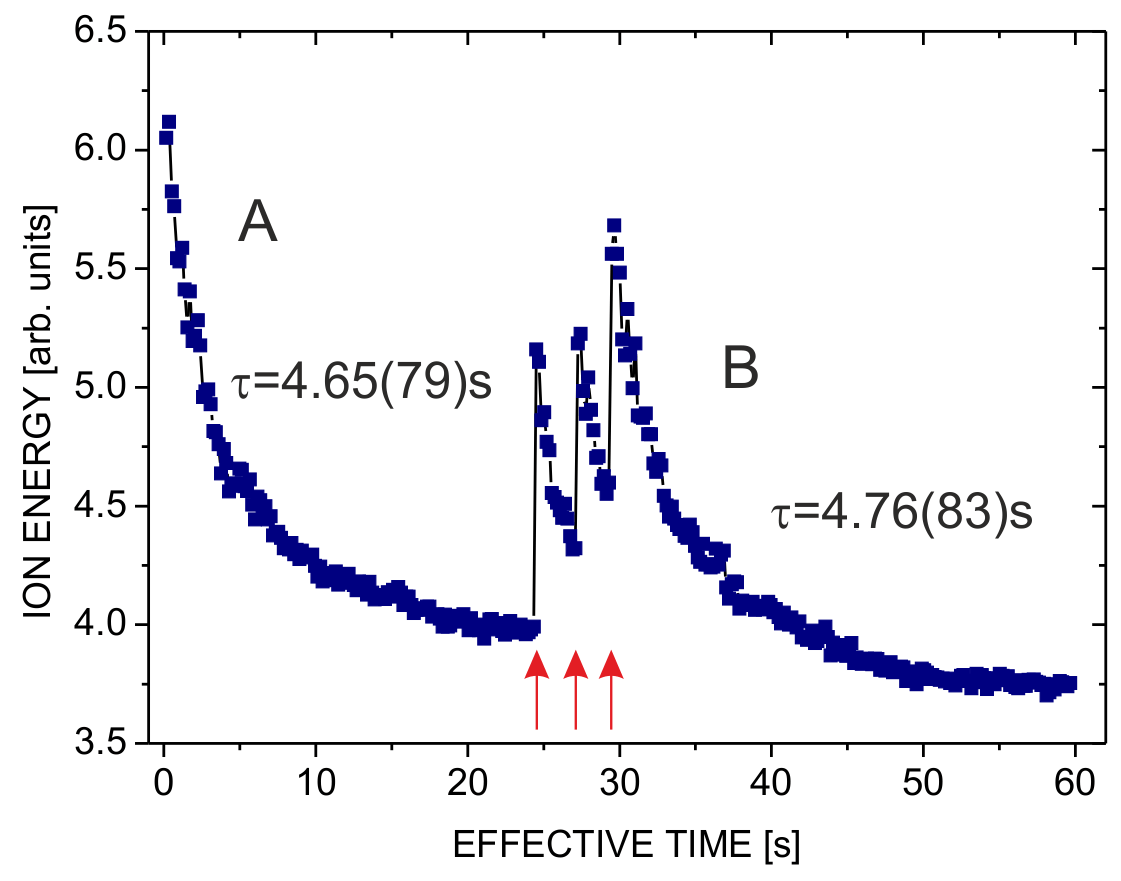}
  \caption{\small (Color online) Ion kinetic energy as a function of time when the ensemble is resistively cooled and re-heated by white noise (indicated by red arrows) several times. The parts marked `A' and `B' are identical to the ones in Fig. \ref{cool1} (top).}
  \label{cycle}
\end{center}
\end{figure}

Figure \ref{cycle} shows
the ion energy as a function of time when the ensemble is cooled and re-heated by white noise several times. In all four cycles which represent different initial energies and trapping times, the observed cooling is exponential and occurs at the same rate. Thus, 'resistive cooling has no memory', as known for the single-particle case, and also as expected from theory for large ensembles. From the production parameters and the space-charge shift, the ion number density is estimated to be around 3 x $10^3$ cm$^{-3}$ in these measurements \cite{sad}.

\begin{figure}[h!]
\begin{center}
  \includegraphics[width=\columnwidth]{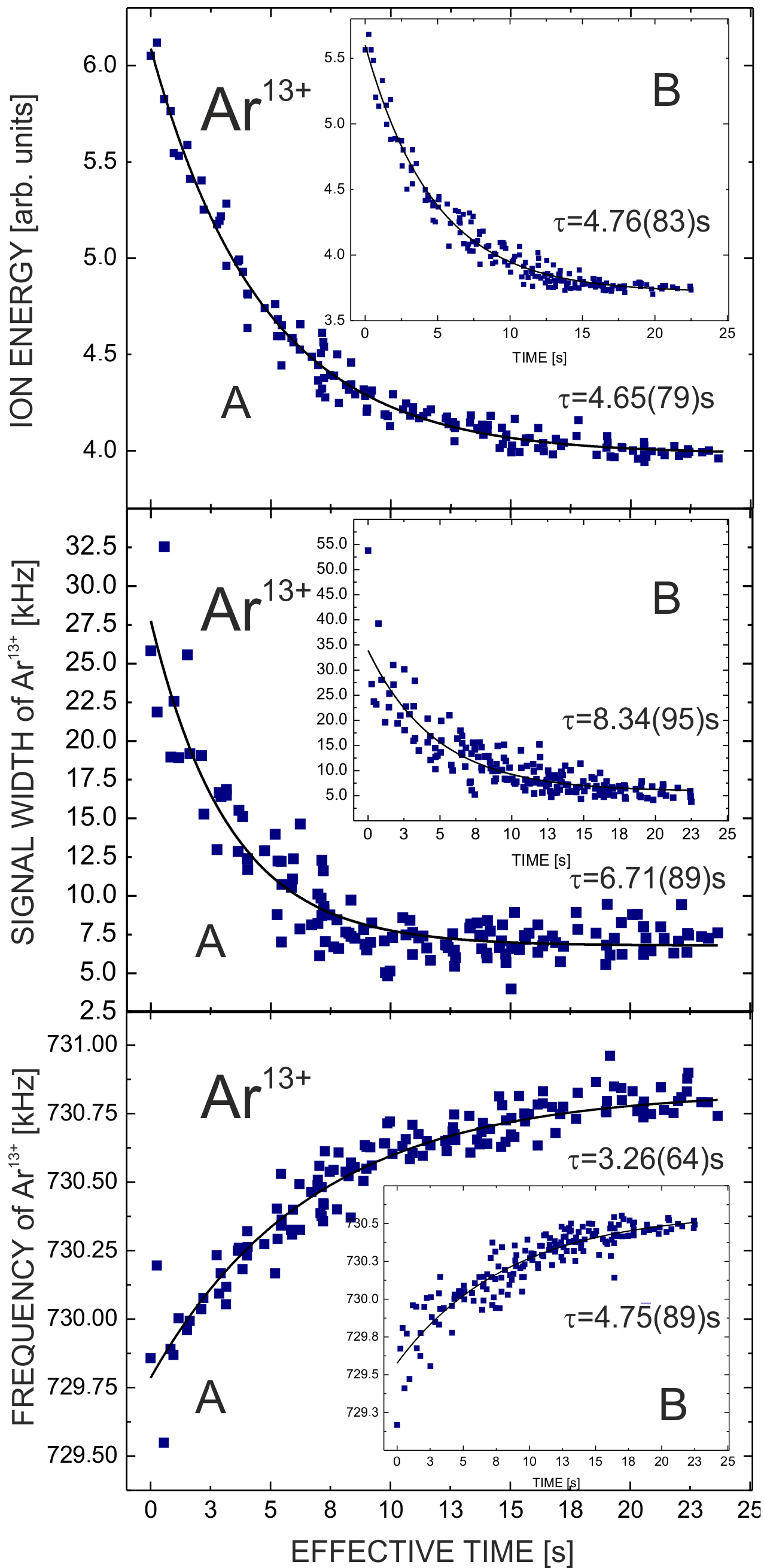}
  \caption{\small (Color online) Effects of resistive cooling as a function of time: on the total ion kinetic energy (top), on the individual thermal signal width of Ar$^{13+}$ (middle), and on the individual center frequency of Ar$^{13+}$.}
  \label{cool1}
\end{center}
\end{figure}

The two indicated parts `A' and `B' of the cooling curve from Fig. \ref{cycle} have been evaluated separately.
In Fig. \ref{cool1} (top), the total ion kinetic energy of measurements A and B is given as a function of the effective time $t$. A single exponential decay of the form
\begin{equation}
E(t)=E_\infty + E_0 e^{-t/\tau}
\end{equation}
fits the data well, yielding a cooling time constant of $\tau=1/\gamma=4.65(79)$\,s (measurement `A'). Here, $E_\infty$ is the energy offset, i.e. the energy at the end point of cooling. A second measurement starting at lower initial energy yields a time constant of $\tau=4.76(83)$\,s (measurement `B'), in agreement with the first one. 

\begin{figure}[h!]
\begin{center}
  \includegraphics[width=\columnwidth]{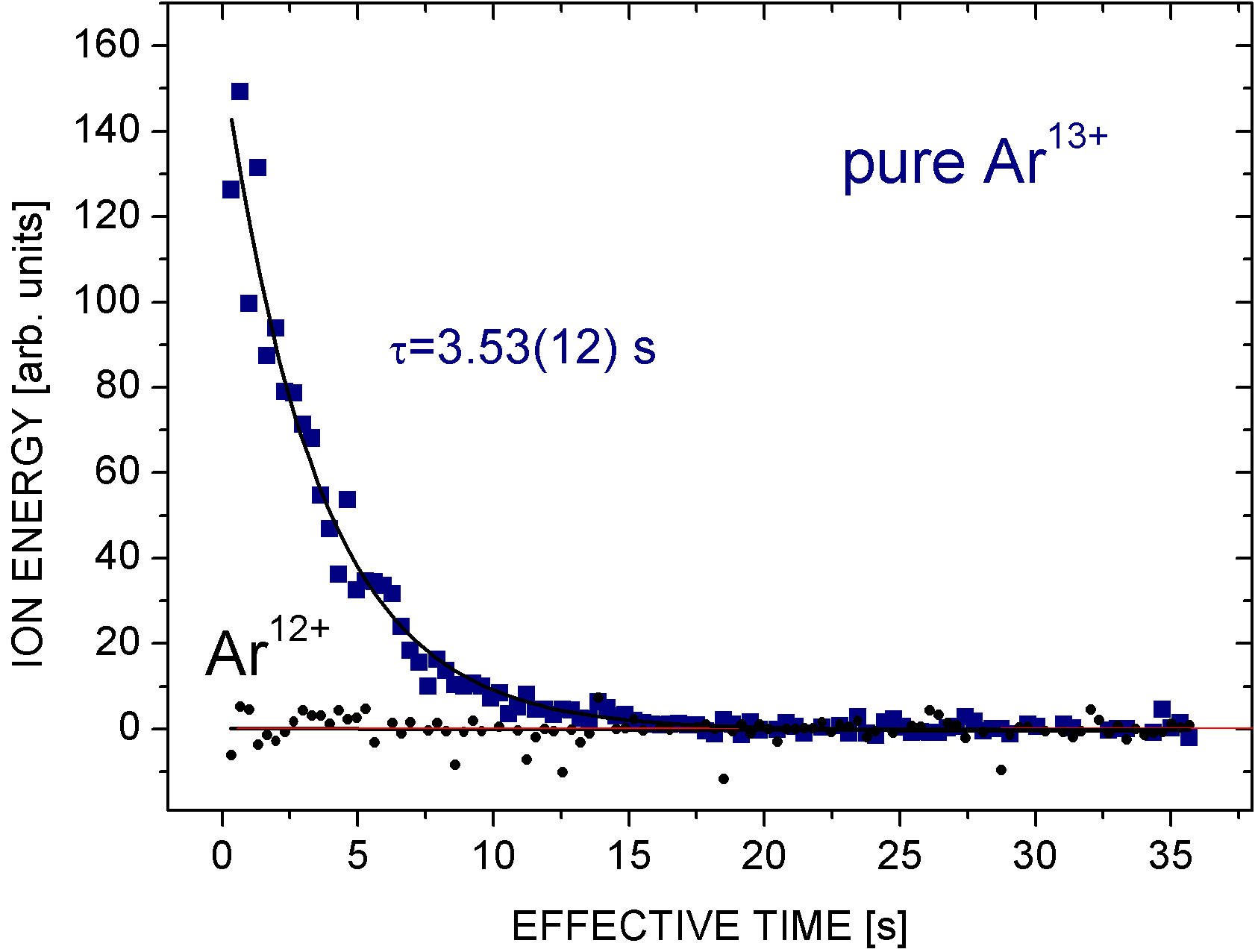}
  \caption{\small (Color online) Ion energy of a pure Ar$^{13+}$ ensemble as a function of time and the corresponding Ar$^{12}$ signal.}
  \label{ar13}
\end{center}
\end{figure}

A similar measurement has been performed with an ensemble of pure Ar$^{13+}$ ions. To that end, the same creation procedure as before has been used, and all unwanted ion species have been removed from the spectrum by selective excitation of their axial motion beyond the confining well of the trap. This has been achieved with the so-called SWIFT technique applied to axial dipole excitation. Fig. \ref{ar13} shows the ion energy as a function of time for Ar$^{13+}$ and the signal for Ar$^{12+}$.
The Ar$^{13+}$ signal shows an exponential decay like observed in the cooling of mixed ion ensembles, and the cooling time constant is in fair agreement with the previous values. The fact that the signal of Ar$^{12+}$ remains at zero for the complete measurement time confirms the efficiency of the SWIFT process and corroborates the above-stated absence of charge exchange which would produce Ar$^{12+}$ from Ar$^{13+}$. 

Let us compare the observed cooling times with expected values. The resonance circuit used for cooling and detection of the axial ion motion is resonant at a frequency of 2$\pi \times 741.23$\,kHz and has a Q-value of about 900. Its inductance is about 3\,mH, and correspondingly it represents a resonance resistance of
\begin{equation}
R=QL\omega_z \approx 12.5 \mbox{M$\Omega$}.
\end{equation}
Inserting this value into Eq. (\ref{cool}) and averaging the effective electrode distance $D$ over the relevant trap volume we have $\langle D^2 \rangle ^{1/2} \approx 40$\,mm, and using the average squared ion charge state $\langle q^2 \rangle$ for q=8 to q=15, we have an expected cooling time constant of $\tau =1/\gamma \approx 4$\,s, which is in agreement with the measured values, both for the mixed and pure ion ensembles.

The observed cooling behavior is different from the one reported in \cite{rs3} as it does not display two largely different cooling time constants with a plateau-like structure between the two domains of the cooling curve belonging to different cooling time constants. Such a feature can apparently only be expected when the axial and radial degrees of freedom are cooled with significantly different time constants, as is the case for the dilute ion cloud in the highly tuned trap used in \cite{rs3}, leading to a complicated time dependence of the observed ion energy \cite{rs3}. In the present work, thermalization amongst all degrees of freedom is made to be faster than resistive cooling at any time during the process, leading to the energies in all degrees of freedom being reduced at the same rate, and hence leading to a single-exponential cooling behaviour.

\subsection{Cooling Effect on Signal Width and Shift}
The middle and lower panels of Fig. \ref{cool1} show the observed signal peak width of Ar$^{13+}$ and the corresponding center frequency of that peak as a function of time, respectively. From our discussion in section \ref{axtrans} we expect the thermal width of each peak to decrease with decreasing energy according to Eq. (\ref{wid}), i.e. to decrease with ongoing cooling time. This is in fact what is observed. The two measured time constants of $\tau=8.34(95)$\,s (for measurement `A') and $\tau=6.71(89)$\,s for measurement `B' agree within their uncertainties. They differ from the time constants observed for the decrease of the ion kinetic energy, because of the higher-order contributions of the temperature in Eq. (\ref{wid}). The presence of higher-order contributions makes the effect of the temperature decrease during cooling non-proportional to the temperature itself, and leads to a different observed time constant. 

The same is true for the shift of the center frequency as a function of time. Also here, the two measurements `A' and `B' agree with each other, but do not necessarily give the same time constant as the ion energy. The observed shift of the center frequency is of the order of several hundreds of hertz which is a relative value of about $10^{-3}$ and can be attributed to the thermal change of the distribution in the presence of the given field imperfections.

\subsection{Final Temperature and Plasma Parameter}
From the known coefficients $C_2$, $C_4$ and $C_6$ of the trap potential and the measured width of the signal peaks, the ensemble temperature can be derived by inversion of Eq. (\ref{wid}), see also Fig. \ref{width}. The final width of the Ar$^{13+}$ signal at the endpoint of cooling is $\delta \omega_z \approx 2 \pi \times 7.3$\,kHz (Fig. \ref{cool1}, middle), which corresponds to a relative width of $\delta \omega_z/\omega_z$ of just below 0.01. According to Eq. (\ref{wid}) and Fig. \ref{width}, this corresponds to a temperature of about 7.5\,K, just above the ambient temperature of 4.2\,K.

According to Fig. \ref{gamma}, resistive cooling to a temperature of 7.5\,K in combination with an ion number density of several 10$^4$ cm$^{-3}$ brings the ion ensembles into the regime of non-negligible correlation with values of $\Gamma$ above 2, where fluid-like behavior can be expected. With our upper limit of ion number densities of several 10$^6$ cm$^{-3}$, values of $\Gamma$ up to 10 are accessible for this temperature. In addition, at these ion number densities, temperatures as high as about 15\,K should still allow us to reach $\Gamma$=2. Note, that the final temperature of resistive cooling is the electronic noise temperature of the circuit, not the physical ambient temperature, as has experimentally been shown in \cite{dje}. Often, the noise temperature is slightly above the ambient temperature, see for example \cite{haff}. Presently, the lowest observed temperature, irrespective of the ion number density and species, is around 7.5\,K.

\subsection{Indication of a Phase Transition}
\begin{figure}[h!]
\begin{center}
  \includegraphics[width=\columnwidth]{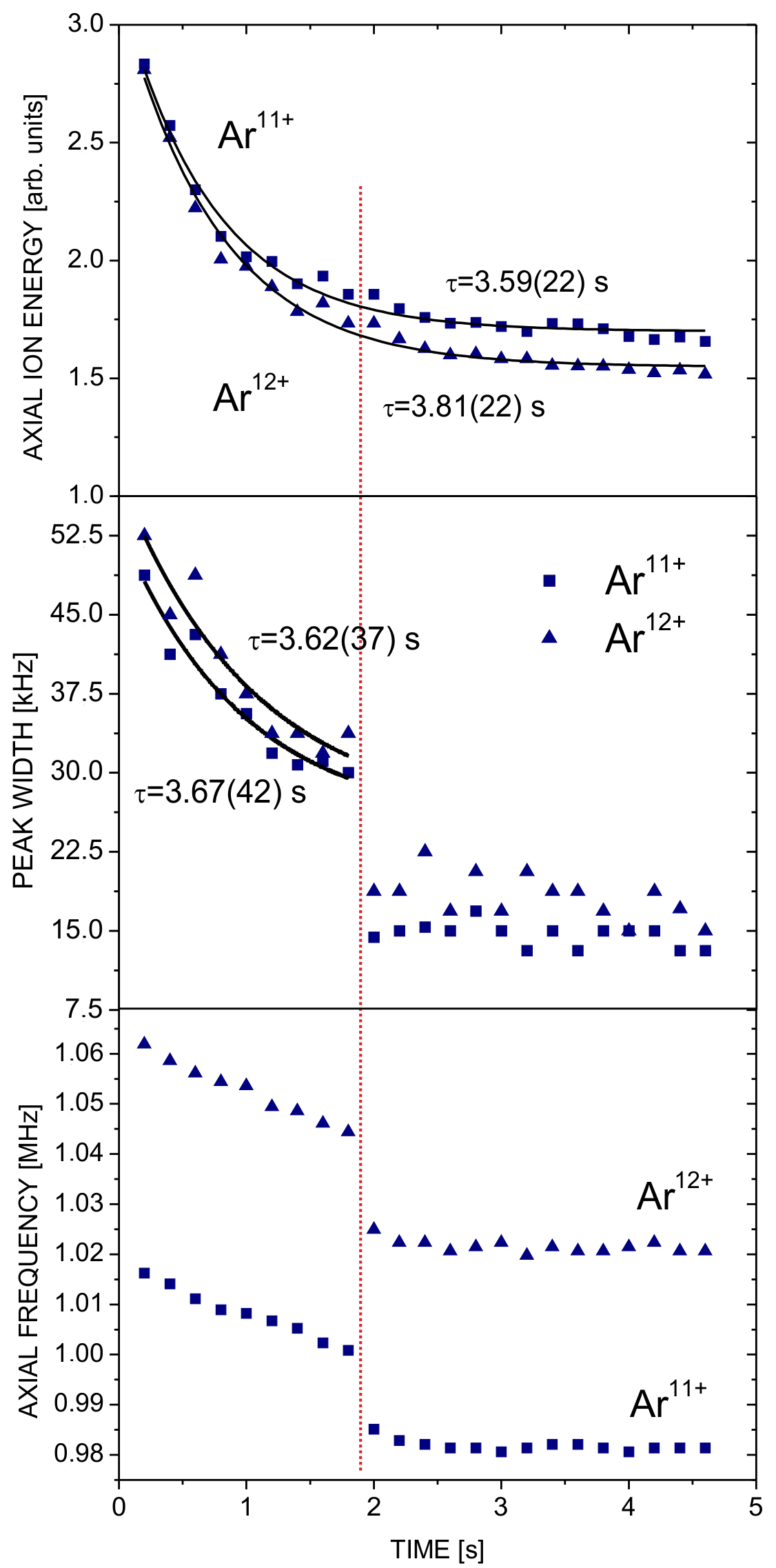}
  \caption{\small (Color online) Measurement as the one shown in Fig. \ref{cool1}, but at much higher ion number density: Effects of resistive cooling on the total ion energy (top), the peak widths (middle), and the axial oscillation frequency (bottom) of Ar$^{11+}$ and Ar$^{12+}$.}
  \label{cool2}
\end{center}
\end{figure}

We have performed cooling measurements like the one shown in Fig. \ref{cool1} at a significantly higher ion number density of a few times $10^6$ cm$^{-3}$. 
Note, that the values of the axial frequencies are different from the previous measurements (Fig. \ref{cool1}) due to different choices of the trapping voltage and the RLC-circuit resonance frequency.
Figure \ref{cool2} shows the effect of resistive cooling on the ion energy, the signal peak width, and the signal center frequency for two selected ion species from the spectrum. 

The axial energy of the ions follows an exponential decay like in all cases discussed above with similar time constants.
The same is initially true for the peak width.
There is, however, a significant discontinuity in the peak width as well as the center frequency of the ion resonances. After that discontinuity, they both remain nearly constant at a low level. The initial behavior of the ion frequency is qualitatively different from the case shown in Fig. \ref{cool1} (bottom): Instead of an exponential dependence with slightly increasing frequency, the frequency here decreases non-exponentially by some kHz. This variation is on the percent scale and hence more pronounced than the thermal shift shown in Fig. \ref{cool1} by more than one order of magnitude. 

The same behavior is also present for the remaining ions in the spectrum (Ar$^{8+}$ to Ar$^{14+}$) and is independent of the step size of the voltage scan, i.e. the discontinuity occurs at the same effective cooling time for all ion species and step sizes. 
The fact that the discontinuity is observed for all ions at the same time during cooling corroborates the assumption of a well-thermalized mixed ion ensemble. 

From the observed relative spectral width $\delta \omega_z/\omega_z \approx 0.035$ at the discontinuity (dotted line in Fig. \ref{cool2}), Eq. (\ref{wid}) and Fig. \ref{width} yield an ensemble temperature of about 14\,K. 
Although the phase transition around 14\,K
collapses the line width to an apparently constant value and therefore it cannot be used to extract the plasma temperature anymore, we believe that the ensemble keeps cooling to a final temperature of around 7.5 K, similar
to the cases where the number density is not sufficient to observe the
phase transition.
While for the measurement in Fig. \ref{cool1} the plasma parameter according to Eq. (\ref{wid}) remains well below $\Gamma = 2$ all of the time, the estimated value at the present discontinuity is $\Gamma \approx 5$. Hence, during cooling, the present ensemble crosses the expected transition from gas-like to fluid-like behavior. 

Based on this, the observed discontinuity can be attributed to the onset of a significant level of correlations of the axial ion motions, leading to a sudden narrowing of the distribution of axial oscillation frequencies (see Fig. \ref{cool2} (middle)) and a sudden downward shift of the axial oscillation frequency by about 25\,kHz out of about 1\,MHz (see Fig. \ref{cool2} (bottom)). 
This can be explained by a higher ion number density in the axial direction leading to a corresponding space-charge shift towards lower axial frequencies. So while the axial ion energies and thus axial oscillation amplitudes of all ions decrease continuously, at the critical temperature the motional correlation changes the oscillation frequency distribution non-continuously.
Based on the observed discontinuities in the ion spectra and the quantitative determination of values of the plasma parameter $\Gamma \approx 5$, we infer that a fluid-like state of the confined ion ensemble has been reached by pure resistive cooling.

\section{Conclusion}
We have measured the resistive cooling behavior of large ensembles of highly charged argon ions in a Penning trap under conditions ensuring efficient thermalization of the ensemble during the cooling process. Under these conditions, the cooling of an ion ensemble is observed as a decrease of the ion kinetic energy that leads to a narrowing and shift of the ion signal peaks. We have shown that all of these temporal evolutions can be described by a simple exponential decay, in contrast to earlier observations made under conditions which lead to inefficient thermalization and consequently to the occurrence of more involved cooling dynamics. 

We have also shown that the exponential nature of resistive cooling of large ion ensembles and their cooling time constant $\tau$ and final temperature are independent of the initial ion kinetic energy and of the number of previous cooling or heating cycles (`resistive cooling has no memory'), as expected from the case of a single ion. Our observations further corroborate the theoretical expectation that the cooling time constant and final temperature of a thermalized ensemble is identical to the single-ion case, which at present is shown to be true for both mixed and pure ion ensembles. 

The final temperatures reached in the present experiments are slightly above the ambient temperature of liquid helium, for which plasma parameters $\Gamma$ up to about 10 are accessible, depending on the actual ion number density. 

While for low ion number densities, the observed cooling behavior is continuous down to the final temperature, for ion ensembles of high density, we have observed a significantly discontinuous behavior of spectral features during cooling, indicating an evolution of the ion ensemble across the $\Gamma \approx 2$ phase transition. From this, we infer that the present ion ensembles reach a fluid-like state at temperatures of around 14\,K without the need to apply laser cooling. 
As a consequence, candidate ions for the envisaged optical spectroscopy measurements \cite{art0} such as Pb$^{81+}$, Bi$^{82+}$, or U$^{91+}$ are expected to form ion crystals by mere resistive cooling, which facilitates highly precise optical and microwave spectroscopy.

\end{document}